\documentclass[epj]{webofc}
\usepackage[varg]{txfonts}   
\usepackage{epstopdf}
\wocname{EPJ Web of Conferences}
\woctitle{INPC 2013}
\begin{document}
\title{Probing the neutron skin thickness in collective modes of excitation}
%
%

\author{N. Paar\inst{1}\fnsep\thanks{\email{npaar@phy.hr}}, 
        A. Horvat  \inst{1}
}

\institute{Physics Department, Faculty of Science, University of Zagreb, Croatia
          }

\abstract{
 Nuclear collective motion provides valuable constraint on the size of neutron-skin thickness and 
 the properties of nuclear matter symmetry energy. By employing relativistic nuclear energy density
 functional (RNEDF) and covariance analysis related to $\chi^2$ fitting of the model parameters, relevant
 observables are identified for dipole excitations, which strongly correlate with the neutron-skin thickness $(r_{np})$, symmetry energy at saturation density $(J)$ and slope of the symmetry energy $(L)$. Using the RNEDF framework and experimental data on pygmy dipole strength ($^{68}$Ni, $^{132}$Sn, $^{208}$Pb) and dipole polarizability ($^{208}$Pb), it is shown how the values of $J$, and $L$, and $r_{np}$ are constrained. The isotopic dependence of moments associated to dipole excitations in $^{116-136}$Sn shows that the low-energy dipole strength and polarizability in neutron-rich nuclei display strong sensitivity to the symmetry energy parameter $J$, more pronounced than in isotopes with moderate neutron-to-proton number ratios. 
}
\maketitle
\section{Introduction}
\label{intro}

Collective motion in neutron-rich nuclei provides the insight into the nuclear structure properties
evolving due to asymmetry in the proton-to-neutron number~\cite{PVKC.07}.
With the advent of new accelerators, exotic nuclei far from the valley of $\beta $ stability have become experimentally accessible. On the neutron rich side, interesting new phenomena arise such as the neutron skin and low-lying dipole transitions known as pygmy dipole strength (PDS). Over
the past years tremendous progress has been achieved in measurements of low-energy dipole transitions~\cite{Sav.13}.
Theoretical frameworks based on nuclear energy density functional,
in conjunction with recent experimental data on pygmy dipole strength~\cite{Kli.07}, dipole polarizability~\cite{Pie.12}, quadrupole excitations~\cite{XRM.13} and charge-exchange excitations~\cite{PVKC.07,Kras.13},
provide complementary constraints of the neutron-skin thickness $r_{np}$.  This quantity is directly connected to the properties of nuclear equation of state, i.e. the
symmetry energy that is of paramount interest for nuclear structure, nuclear reactions and astrophysics. In order to identify which observables 
are useful to determine specific property of nuclei or nuclear matter, recently it has been shown that covariance analysis in connection to $\chi^2$ 
fitting of the energy density functional parameters can provide the estimates of correlations between various quantities of interest~\cite{Reinh.10}.
Especially interesting are correlations between the observables related to excitation phenomena in finite nuclei and the symmetry energy at saturation density $(J)$, slope of the symmetry energy $(L)$~\cite{Kli.07}, and $r_{np}$. The neutron-skin thickness, representing an isovector quantity, displays strong linear correlation with the symmetry energy parameters $J$ and $L$~\cite{Fur.02,RM.11}.
By employing the relativistic nuclear energy density functional, in this article we explore the underlying structure of low-energy dipole transition strength, and its relations with the symmetry energy parameters ($J$,$L$) and $r_{np}$ in finite nuclei. Relevant correlations between various quantities are studied using covariance analysis in the framework of the relativistic nuclear energy density functional.

\section{Pygmy dipole strength in neutron rich nuclei}

\begin{figure}
\centering
\includegraphics[width=8 cm,clip]{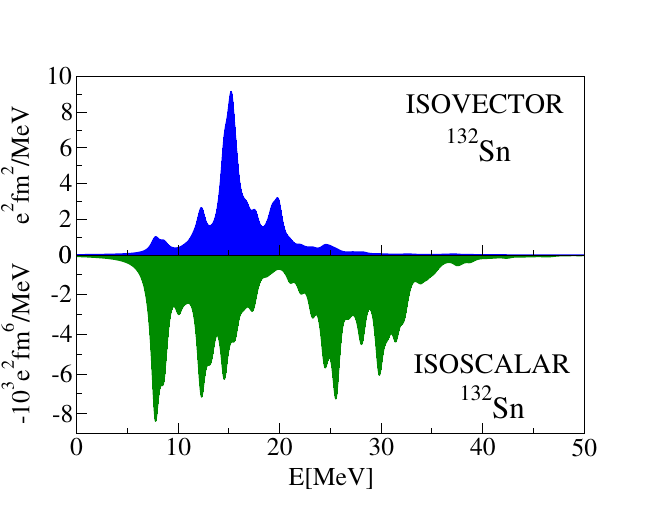}
\caption{Isovector (upper panel) and isoscalar (lower panel) dipole transition strength distributions for $^{132}$Sn, calculated using RQRPA with DD-ME2 effective interaction. Discrete transition spectra are folded by a Lorentzian of a fixed width $\Gamma = $1 MeV.}
\label{IS.IV}       
\end{figure}

\begin{figure}
\centering
\includegraphics[width=7.5 cm,clip]{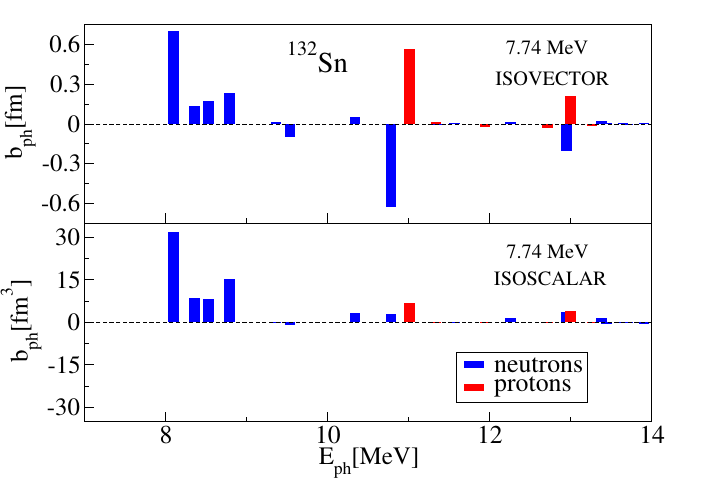}
\caption{Comparison between partial contributions of $ph$ transitions to the overall isovector and isoscalar transition strength for the state at 7.74 MeV in $^{132}$Sn.}
\label{IS.E2qp}       
\end{figure}

Model calculations of dipole transitions in nuclei based on the relativistic nuclear energy density functional (RNEDF) are realized in terms of relativistic Hartree-Bogoliubov (RHB) model and relativistic quasiparticle random phase approximation (RQRPA)~\cite{PVKC.07}.
In the past decade this microscopic approach reached an adequate level of accuracy in comparison with experimental data. 
The nuclear response to an external excitation is described with the reduced transition probability for the corresponding transition operators~\cite{PVKC.07}. Figure \ref{IS.IV} shows the transition strength distributions for $^{132}$Sn using the isovector and isoscalar dipole transition operators~\cite{PVKC.07}, calculated using the RQRPA with DD-ME2 effective interaction~\cite{Lal.05}. In the case of isovector transitions, the giant resonance peak appears as dominant structure, while the pygmy dipole strength below 10 MeV exhausts only a small fraction of the overall transition strength. In comparison to the isoscalar case (Fig.~\ref{IS.IV}), the difference is rather evident, but an interesting feature appears in the low-energy part. The states of the strongest
low-energy (E$<$10 MeV) transition strength appear exactly at the same excitation energy both in the isoscalar and isovector channel. In order to explore the nature of these low-energy isovector and isoscalar transitions, we study the contributions of individual two-quasiparticle $(2qp)$ configurations to the total transition strength for excited state of given energy, $B^T(E1, \omega_\nu)$.  We examine which $2qp$ (or particle-hole) pairs  contribute to a certain QRPA state in order to assess the collectivity of the state~\cite{Vre.12}. The relative sign of the contribution is also an important factor. In terms of particular $2qp$ contributions, the overall transition strength is given by
\begin{equation}
B^{T}(E1, \omega_\nu)=\left|\sum_{\kappa, \kappa'} b_{\kappa,\kappa'}^ {T,\nu}\right|^2,
\label{strength}
\end{equation}
where $b_{\kappa,\kappa'}^ {T, \nu}$ denotes contribution of $(\kappa, \kappa')$ $2qp$ configuration
to the transition strength.
\begin{figure}
\centering
\includegraphics[width=7.05 cm,clip]{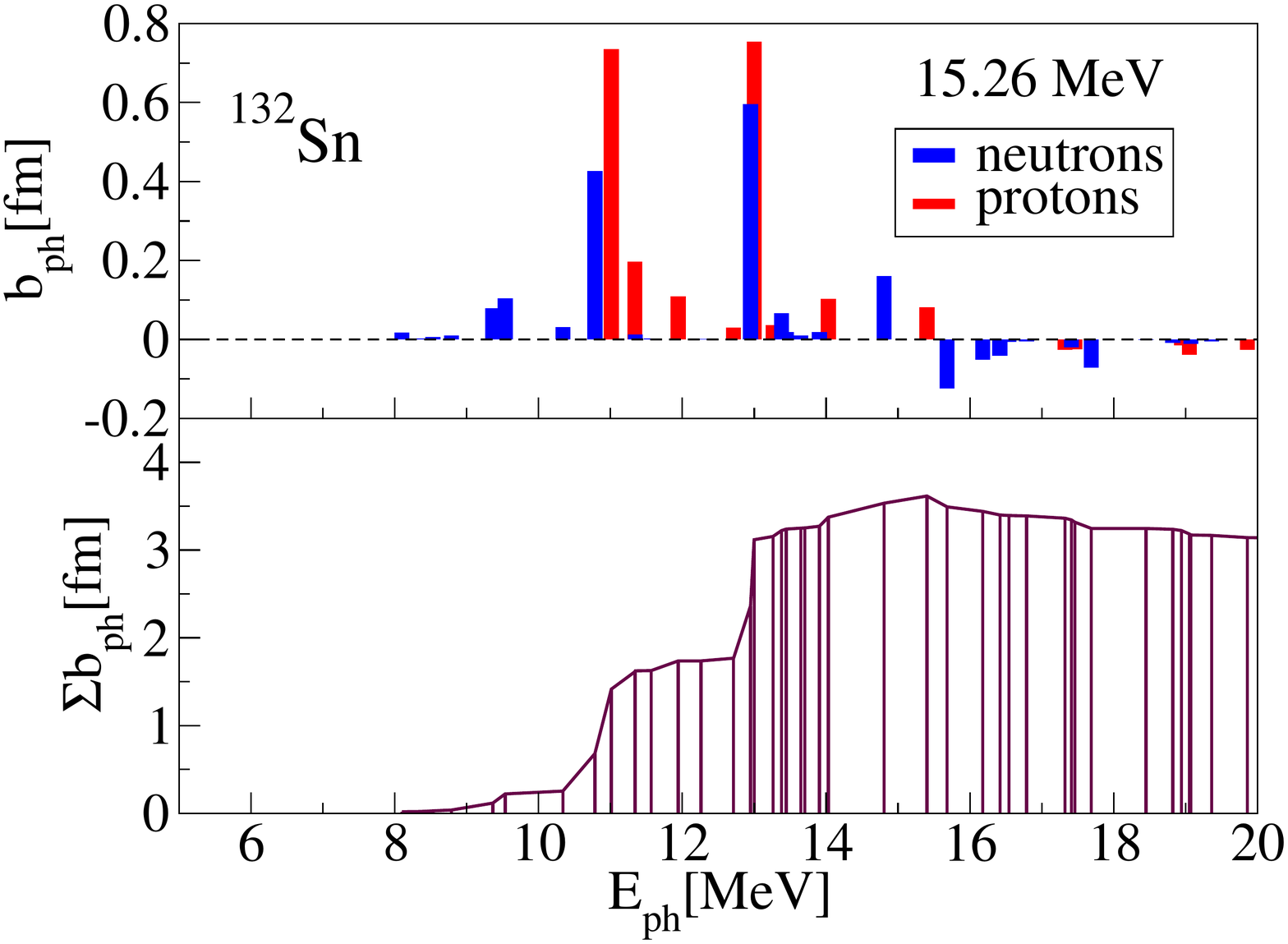}
\includegraphics[width=7.05 cm,clip]{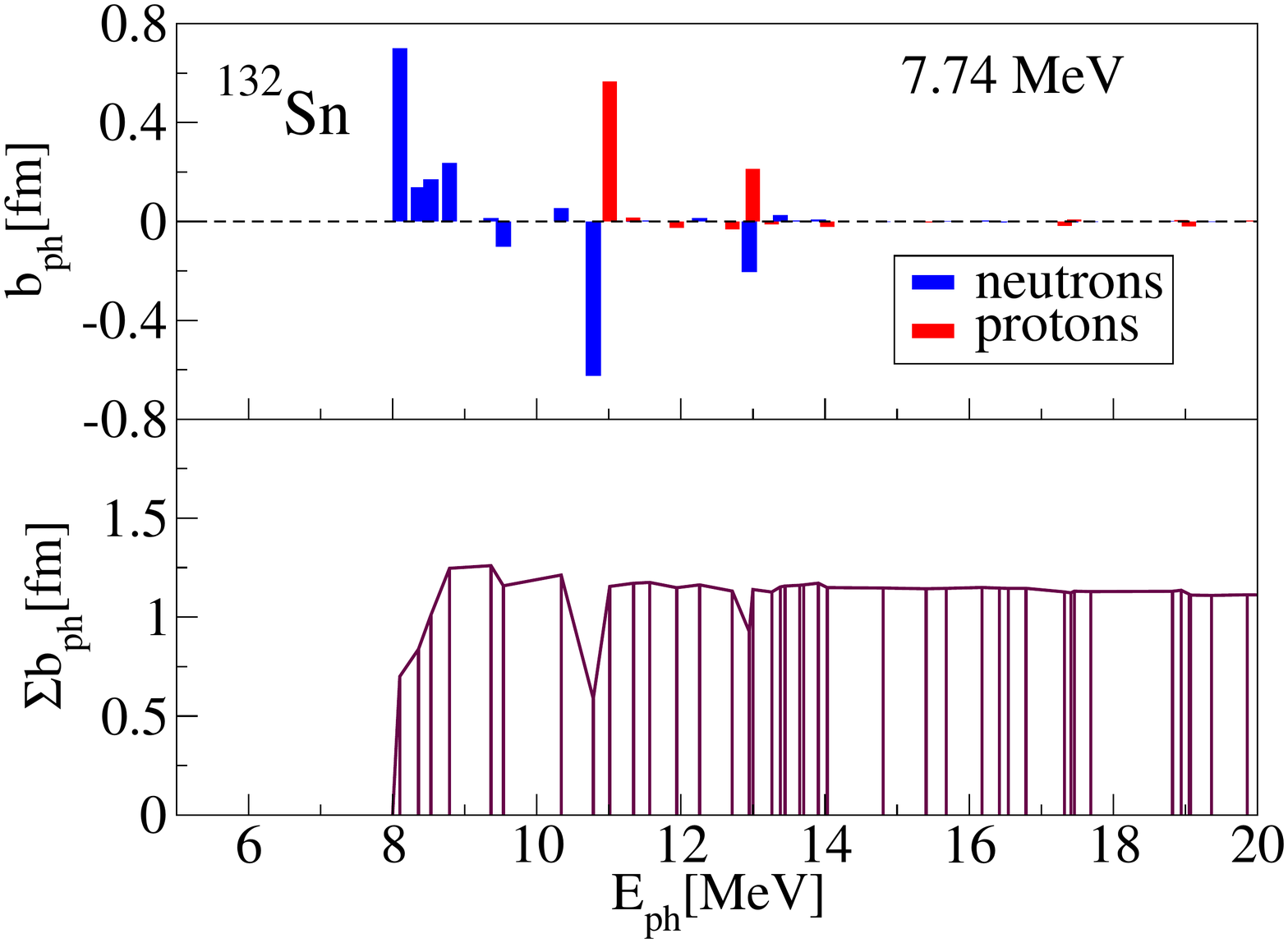}
\caption{(Left side) Contributions of individual $ph$ transitions (with the energy of $ph$ configuration $E_{ph}$) to the overall isovector dipole transition strength (\ref{strength}) for the state at excitation energy 15.26 MeV for $^{132}$Sn; the lower panel displays the running sum of the corresponding transition strength as the energy of $ph$ pairs increases. (Right side) The same as on the left side, but for the low-energy state at energy 7.74 MeV.}
\label{buildup1}       
\end{figure}
Figure \ref{buildup1} shows contributions from the individual $2qp$, i.e., particle-hole ($ph$) configurations to the isovector dipole transition strength of states at 15.26 MeV (GDR state), and 7.74 MeV (PDS state) for $^{132}$Sn. Each $ph$ pair is denoted by its unperturbed excitation energy, $E_{ph} = E_p - E_h$. The respective lower panels in the figure show the running sum of the transition strength with increasing $E_{ph}$ energy.
We notice a high degree of coherence in different proton and neutron contributions to the GDR strength. In the case of PDS, some level of coherence in neutron transitions is obtained, however, we also notice partial cancellation of the transition strength. It is interesting to note that transitions between the same pairs of protons ($1g_{9/2} \rightarrow 1h_{11/2}$ and $1f_{5/2} \rightarrow 1g_{7/2}$) and neutrons ($1g_{7/2} \rightarrow 1h_{9/2}$ and $1h_{11/2} \rightarrow 1j_{13/2}$) with equal $ph$ energies act coherently in the case of GDR and destructively in the case of the low-lying state. The total $B^{T}(E1, \omega_\nu)$ builds up gradually with $E_{ph}$ for the GDR state. In the case of PDS, a coherent buildup by neutron contributions occurs for $E_{ph}<$ 10 MeV and the strength remains almost constant thereafter. We note that it is not possible to uniquely determine which transitions cancel each other only by using information shown in Figure \ref{buildup1}. One could assume that, for instance, neutron transitions at lower $ph$ energies are canceled by those at higher energies. However, the isoscalar low-lying state at energy 7.74 MeV in $^{132}$Sn (Figure \ref{IS.E2qp}), consists almost exclusively of coherent neutron contributions, from exactly the same $ph$ pairs at low $E_{ph}$ as in the case of the isovector excitations at the same RPA energy (Fig.~\ref{IS.E2qp}, Ref.~\cite{Vre.12}). The structure of low-energy dipole states obtained in the relativistic framework~\cite{Vre.12} appears qualitatively consistent with the results based on the Skyrme functional~\cite{Roca.12}.

\section{Constraining the nuclear symmetry energy and neutron-skin thickness from pygmy dipole strength and dipole polarizability}
 
\begin{figure}
\centering
\includegraphics[width=8 cm,clip]{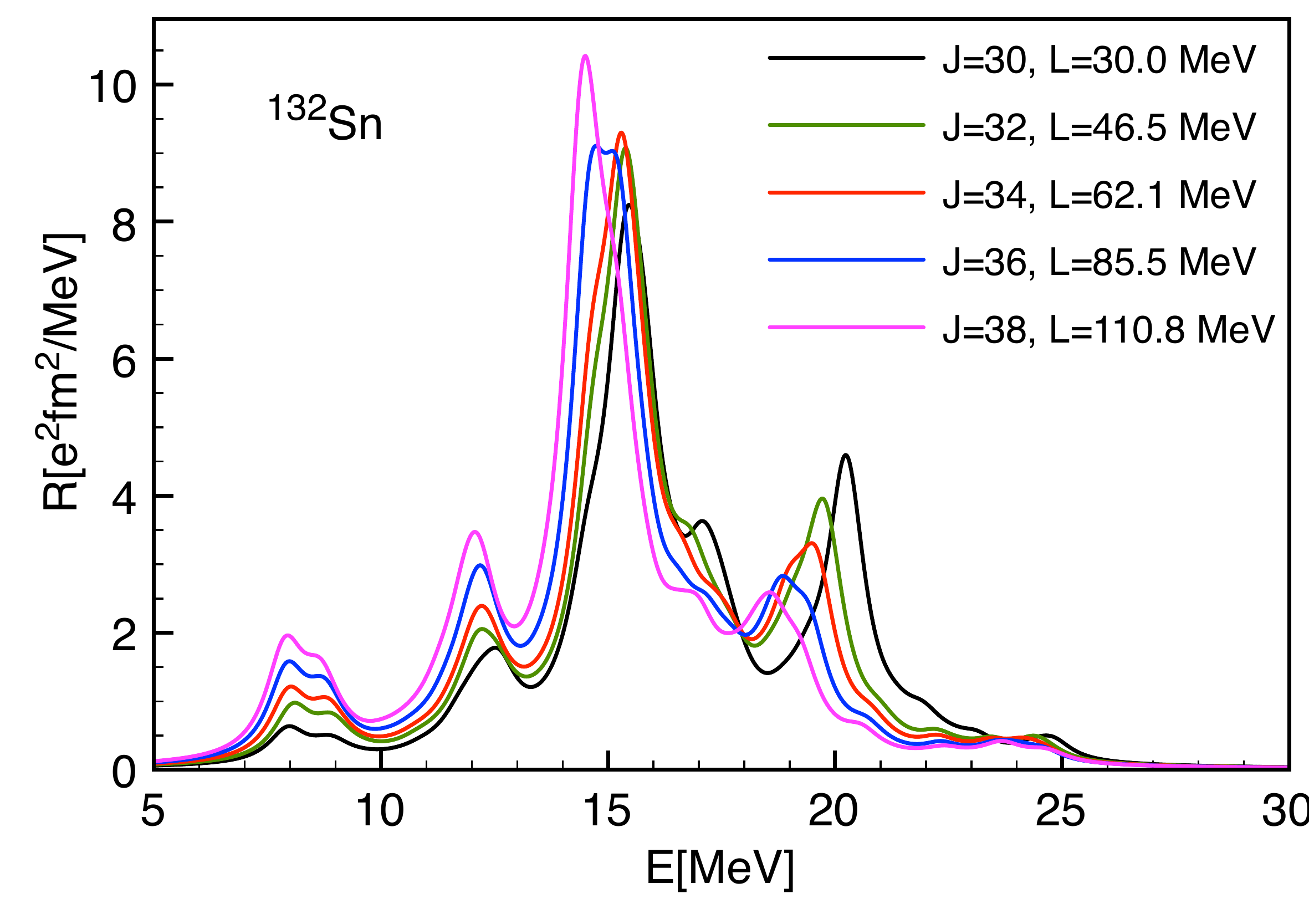}
\caption{Isovector dipole transition strength for $^{132}$Sn for the set of density dependent effective interactions
spanning the range of values $J=$30,32,...,38 MeV, and the corresponding $L$ as denoted in the legend.}
\label{Sn132excstr}       
\end{figure}

The nuclear energy density functional allows to establish a direct link between the symmetry energy at saturation density $(J)$, slope of the symmetry energy $(L)$ and collective modes of excitation by employing the same functional in description of nuclear matter properties and at the level of RPA.
In order to illustrate this connection, we use a set of relativistic density dependent meson-exchange effective interactions (DD-ME), accurately  calibrated on the same set of experimental data, but with an additional constraint on $J$~\cite{Vret.03}. In this way a set of consistent parameterizations is obtained, allowing search for various dependences of the excitation properties on $J$ and $L$.  Figure~\ref{Sn132excstr} shows the isovector dipole transition
strength distributions for $^{132}$Sn, obtained using a set of DD-ME effective interactions spanning the range of values $J$=30-38 MeV (and correspondingly $L$=30-110.8 MeV). Obviously the transition spectra depend sensitively on $J$ and $L$. The peak energy of the GDR systematically
decreases with the increasing $J$ and some sensitivity of the transition strength is also observed. In the low-energy region, the PDS strength
displays strong sensitivity on $J$, i.e. the transition strength considerably increases with $J$ (by factor $\approx$ 3-4), similar as in the previous study by Piekarewicz ~\cite{Piek.11}.
The observables related to dipole excitations can be described in terms of the moments
\begin{equation}
m_k=\sum_{\nu} E_{\nu}^k B(E_{\nu}),
\end{equation}
where $B(E_{\nu})$ denotes the transition strength at excitation energy $E_{\nu}$.
The $k=1$ moment corresponds to the well known TRK sum rule and the $k=-1$ moment is directly proportional
to the dipole polarizability $\alpha_D$, while $m_0$ gives the total strength. Since the low-energy strength weigh
more in the sum $m_{-1}$ than in $m_1$, the PDS exhausting only a few percent of the EWSR
can exhaust more than $20\%$ of the dipole polarizability in neutron rich nuclei \cite{Piek.11,Vre.12}. 

By varying the symmetry energy parameters $J$ and $L$, two effects influence the dipole polarizability; the change in the RPA excitation strength and the shift of the energy peak, considering that each contribution of the transition strength to the total $\alpha_D$ is inverse energy weighted.
Since the increase in $J$ leads to decrease of the GDR excitation energy, and dipole transition strength grows (Fig.~\ref{Sn132excstr}), both effects increase the dipole polarizability, which makes $\alpha_D$ an excellent probe for $J$. Conversely, these two effects act oppositely on $m_1$, making it fairly insensitive to the symmetry energy at saturation in accordance with the TRK sum rule. 
In a recent experiment using high resolution $(p, p')$ scattering the dipole polarizability in $^{208}$Pb has been measured, $\alpha_D=(20.1 \pm 0.6) fm^3$ \cite{Tam.11}. In Ref.~\cite{Pie.12} an elaborate set of both relativistic and non-relativistic nuclear 
energy density functionals (EDFs) have been employed to constrain the neutron skin thickness in $^{208}$Pb from $\alpha_D$.
By averaging over various predictions, estimates are provided with associated systematic errors for $r_{np}$ 
and $\alpha_D$ for $^{48}$Ca, $^{132}$Sn, and $^{208}$Pb. It has been shown that precise measurements 
of $r_{np}$ in both $^{48}$Ca and $^{208}$Pb in combination with the recent measurement of $\alpha_D$ should
significantly constrain the isovector sector of the EDFs. More recently, droplet model has been used as a guide to better understand the correlations between $\alpha_D$
and other isovector observables~\cite{Xavi.13}. By employing a representative set of relativistic 
and nonrelativistic EDFs, it has been shown that actually $\alpha_D J$ 
is far better isovector indicator than $\alpha_D$ alone. 

\begin{figure}
\centering
\includegraphics[width=8 cm,clip]{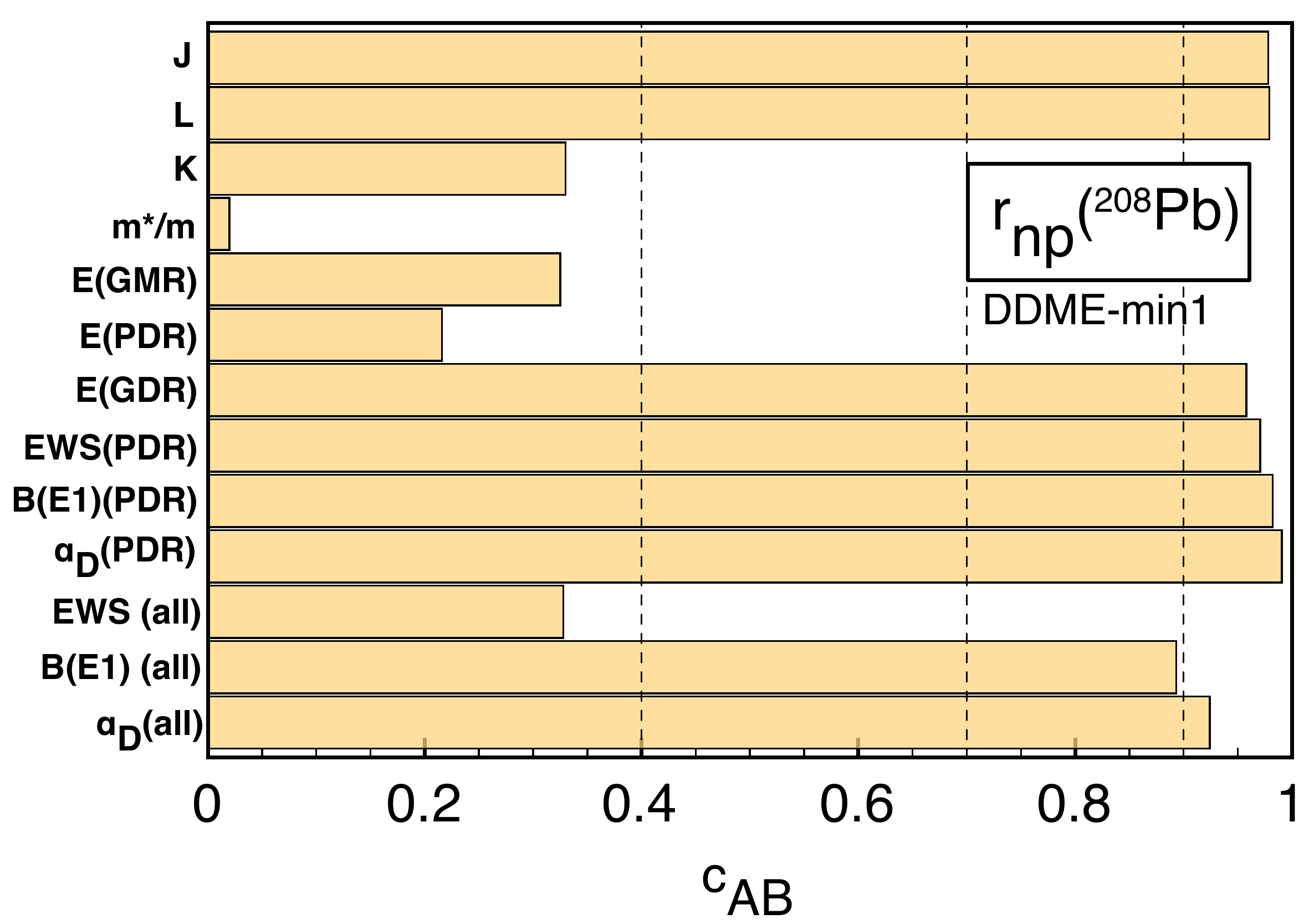}
\caption{Correlations between the neutron-skin thickness in $^{208}$Pb and nuclear matter properties ($J$,$L$,$K$,$m^*/m$); the excitation energies of isoscalar giant monopole resonance (GMR), isovector giant dipole resonance (GDR), and pygmy dipole strength (PDR); energy weighted PDR strength , PDR strength, inverse energy weighted PDR strength ($\alpha_D$); total energy weighted strength, total transition strength and total dipole polarizability.}
\label{correlation}       
\end{figure}

An alternative approach to identify possible correlations between various properties of finite nuclei, nuclear excitations and nuclear matter properties
is covariance analysis on the ground of $\chi^2$ fitting of the nuclear EDF to the experimental data~\cite{Reinh.10}.
The covariance analysis represents powerful tool to assess relevant correlations, given in terms of Pearson correlation coefficients,
defined as the covariance of the two variables divided by the product of their standard deviations. 
The present covariance analysis is based on the relativistic framework
with density-dependent meson-nucleon couplings \cite{Nik.02}. For
the purpose of calculating the relevant correlations, the fitting of the model
parameters has been systematically done by using the set of observables for
17 nuclei from $^{16}$O toward $^{214}$Pb, including binding energies, charge radii,
diffraction radii, and surface thicknesses, assuming the spherical symmetry.
In this way, the parameterization DDME-min1 has been obtained, including
the covariance matrix and Pearson correlation coefficients.
As a representative case of the present correlation analysis, we explore relations between the neutron-skin
thickness in $^{208}$Pb and various properties of nuclear excitations and nuclear matter. Figure~\ref{correlation}
shows the Pearson correlation coefficients for neutron-skin thickness in $^{208}$Pb and (i) symmetry energy at saturation density $J$,
slope of the symmetry energy $L$, nuclear matter incompressibility $K$, effective mass $m^*/m$, (ii) excitation
energies of isoscalar giant monopole resonance (GMR), isovector giant dipole resonance (GDR), low-energy PDS
strength, (iii) moments of the low-energy isovector dipole transitions $m_1$, $m_0$, $m_{-1}$, (iv) moments of the overall
dipole transitions. Since the low-energy strength is clearly separated from the GDR one (similar as in Fig.~\ref{Sn132excstr}), the selection of the energy cut-off between low- and high-energy part of transition spectra is straightforward. In the case of $^{208}$Pb, the energy cut-off is selected at 10 MeV, but we have also checked
that the results of the correlation analysis remain stable also for 11 and 12 MeV.
As expected, the symmetry energy parameters $J$ and $L$ are strongly correlated with $r_{np}$.
The excitation energy of GMR, that is an isoscalar quantity, appears uncorrelated with $r_{np}$. On the other hand, strong indicators of isovector properties
are the moments related to the isovector dipole transitions. At the level of the overall transition strength, the dipole
polarizabiltiy and B(E1) transition strength represent strong isovector indicators. The properties of low-energy 
dipole transitions (all the moments $m_1$, $m_0$, $m_{-1}$) appear strongly correlated with $r_{np}$.
This result is at variance with previous study based on Skyrme functional, and clearly indicates some level
of the model dependence in correlation analysis~\cite{Reinh.10}, that may originate in density dependence of the energy density
functional or in various fitting protocols employed. 

\begin{figure}
\centering
\includegraphics[width=7 cm,clip]{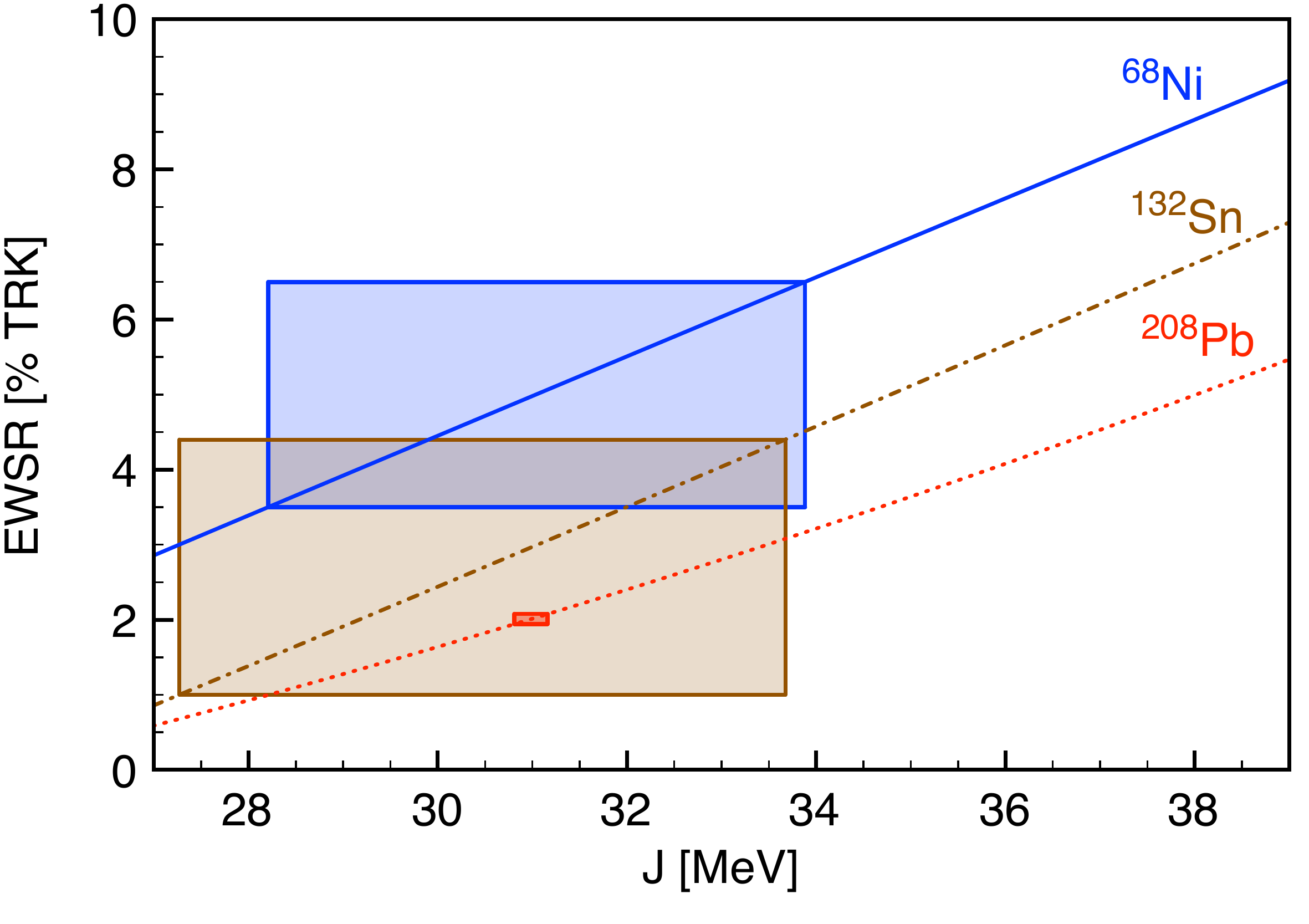}
\includegraphics[width=7 cm,clip]{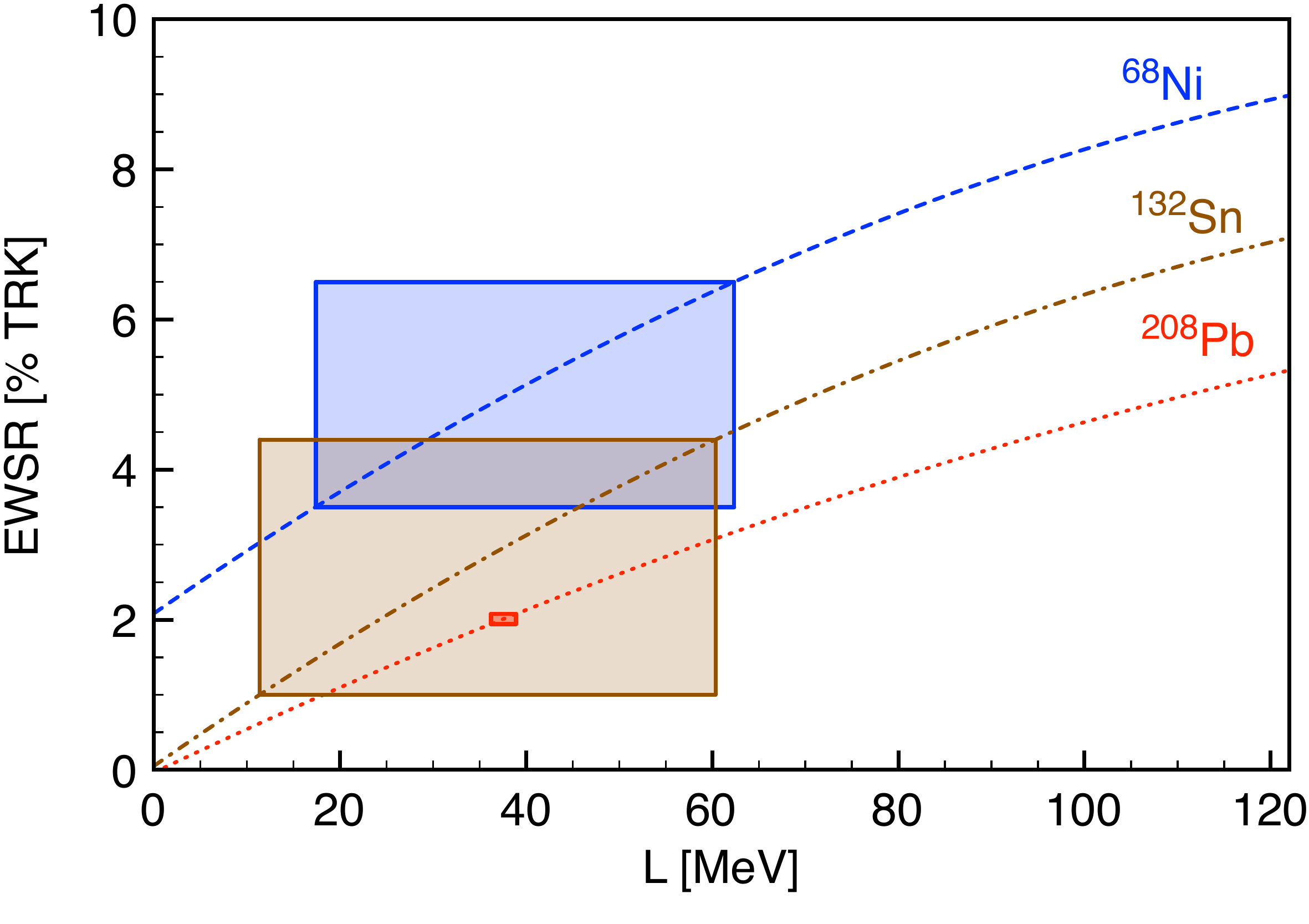}
\caption{Theoretical dependencies of the PDS energy weighted strength with 
respect to the classical TRK sum rule as a function of $J$ (left panel) and $L$ (right panel)
for $^{68}$Ni, $^{132}$Sn and $^{208}$Pb. The corresponding experimental 
data~\cite{Wie.09,Adr.05,Tam.11}
constrain the values of $J$ and $L$ from theoretical results (denoted with 
rectangles). 
}
\label{ewsrpds}       
\end{figure}

Since in the present analysis the PDS strength appears correlated with the symmetry energy 
parameters $J$ and $L$, as well as with
the neutron-skin thickness, the experimental data on PDS can be employed to constrain these quantities.
We employ the experimental data on the PDS energy weighted strength for
$^{68}$Ni~\cite{Wie.09}, $^{132}$Sn~\cite{Adr.05} and $^{208}$Pb~\cite{Tam.11} and model calculations
using the set of relativistic functionals spanning the range of values of the symmetry energy at saturation
density~\cite{Vret.03}. Similar study, but with another selection of
the effective interactions has recently been performed by Carbone et al.~\cite{Car.10}.
Figure~\ref{ewsrpds} shows the results of model calculations for the energy weighted PDS strength 
as functions of $J$ (left panel) and $L$ (right panel). By imprinting the experimental data for the PDS
strengths on theoretical curves, the constraints on $J$ and $L$ are obtained. The results
for $^{68}$Ni and $^{132}$Sn appear in reasonable agreement. Due to
measurement of outstanding accuracy for $^{208}$Pb~\cite{Tam.11}, in this case the values of $J$ and $L$ are 
constrained with narrow uncertainties compared to other nuclei.

\begin{figure}
\centering
\includegraphics[width=14.0 cm,clip]{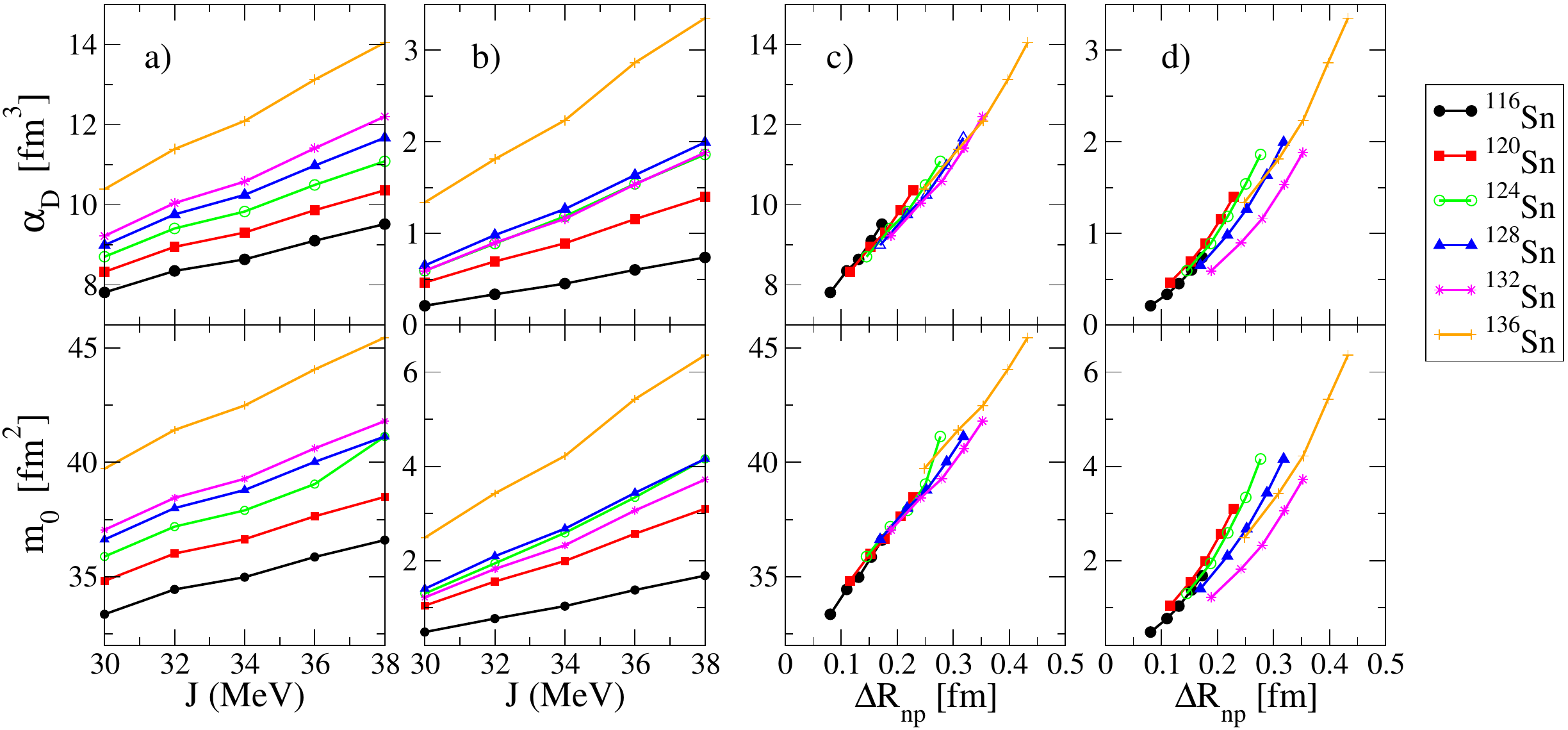}
\caption{Dipole polarizability ($\alpha_D$) and B(E1) transition strength $(m_0)$ for $^{116-136}$Sn isotopes,
including the overall spectra (a),(c), and only PDS transitions (b),(d), displayed
as functions of $J$ and $r_{np}$.}
\label{trendsSn}       
\end{figure}

In view of the future experiments involving neutron-rich nuclei, it is interesting to explore the sensitivity
of the relevant observables on $J$ and $r_{np}$ in systematic variations of the
neutron number. Figure \ref{trendsSn} shows that in $^{116-136}$Sn isotope chain, nuclei
with the largest neutron excess display stronger sensitivity of the low-lying and overall moments to $J$. 
However, this appears not to be the case with $r_{np}$ as a variable, where a certain degree 
of shell effects becomes apparent. A similar effect on the PDS contribution
to the total EWSR for various tin and nickel isotopes was also noticed in Refs.~\cite{Piek.11} and~\cite{Piek.06}, 
where $^{120}$Sn and $^{68}$Ni were found to be better candidates than $^{132}$Sn and $^{78}$Ni, for finding correlations between the neutron skin thickness and $m_1(PDS)$.
It was noted that the PDS share in the total EWSR decreases after $^{120}$Sn, due to the contribution of half-filled $1h_{11/2}$ neutron orbital, which because of its large angular momentum contributes heavily to the neutron skin thickness, but transitions starting from and ending in the $1h_{11/2}$ neutron orbital are high in $2qp$ energy because of parity selection rules. Therefore their influence on the PDS strength 
lies in the decoherent $E_{2qp}$ region.
\section{Conclusion} 

Pygmy dipole strength represents a unique mode of excitation: its strength is governed by coherent low-$E_{2qp}$ neutron configurations
and by decoherence of the proton and neutron $2qp$ configurations which are on the other side relevant for the collectivity of GDR. Decoherence mechanism is responsible for the small strength of the pygmy mode, in contrast to relatively strong transition strength at the same RPA eigenvalue in the isoscalar channel, where the decoherence effect is absent. At the PDS energy, exactly the same neutron configurations with  $E_{2qp} < $~10 MeV participate relatively by the same amount in the buildup of the response both for the isovector and isoscalar dipole transition operators. By employing the DDME-min1 effective interaction, introduced for statistical covariance analysis, it is shown that relevant correlations involving the neutron-skin thickness and symmetry energy parameters can be identified. The dipole strength-related observables (PDS moments, overall dipole polarizability and transition
strength) are strongly correlated with $J$ and $L$ parameters of the symmetry energy, and the neutron skin thickness.
New experimental data on PDS transitions and dipole polarizability, together with the model calculations based on nuclear energy density functionals, enable constraints on the density dependence of the symmetry energy and neutron skin thickness. It is shown that the PDS in nuclei with large neutron excess provides improved constraints for $J$. Charge-exchange excitations in neutron-rich nuclei provide another feasible 
approach to constrain $r_{np}$,  $J$, and $L$. Recently it has been shown that the excitation energies
of the anti-analog giant dipole resonance represent new stringent constraint on the value of
neutron skin thickness~\cite{Kras.13}.

%

\end{document}